\documentclass{PoS}

\usepackage[english]{babel}

\usepackage[utf8]{inputenc}

\usepackage{xcolor}

\usepackage{mathtools}

\usepackage{booktabs}

\usepackage{cite}

\newcommand{\mr}[1]{\mathrm{#1}}

\title{$Z_\mathrm{S}/Z_\mathrm{P}$ from three-flavour lattice QCD}

\ShortTitle{$Z_\mathrm{S}/Z_\mathrm{P}$ from three-flavour lattice QCD}

\author{Jochen Heitger and \speaker{Fabian Joswig} 	\hfill\parbox{17.5mm}{\vspace{-0.25cm}\raggedleft\footnotesize\it%
		MS-TP-18-22
	}
	\\
        Westfälische Wilhelms-Universität Münster, Institut für Theoretische Physik, \\ Wilhelm-Klemm-Straße 9, 48149 Münster, Germany \\
        E-mail: \email{fabian.joswig@wwu.de}}

\author{Anastassios Vladikas\\
        INFN, Sezione di Tor Vergata, c/o Dipartimento di Fisica, Università di Roma Tor Vergata, \\
        Via della Ricerca Scientifica 1, 00133 Rome, Italy
        }

\abstract{We report on advances in the non-perturbative determination of the ratio $Z_\mathrm{S}/Z_\mathrm{P}$ of the pseudoscalar to the scalar renormalization constants in three-flavour lattice QCD with Wilson-clover quarks and tree-level Symanzik improved gluons. The computations are based on the Ward identity approach, using Schrödinger functional boundary conditions. Our results for $Z_\mathrm{S}/Z_\mathrm{P}$ cover a range of couplings along a line of constant physics with lattice spacings of about $0.09\,$fm and below, relevant for phenomenological applications such as the calculation of renormalized quark masses.}

\FullConference{The 36th Annual International Symposium on Lattice Field Theory - LATTICE2018\\
		22-28 July, 2018\\
		Michigan State University, East Lansing, Michigan, USA.}

\begin{document}

\section{Introduction}
Quark masses are key parameters of the standard model of particle physics, which are not directly accessible by experiment. Lattice QCD offers a non-perturbative first-principles method to calculate these masses with controlled uncertainties. In the work presented here we use the $\mathrm{O}(a)$ improved Wilson formulation for the quarks in three-flavour QCD, which has the disadvantage of explicit chiral symmetry breaking. As a direct consequence, the scale dependent renormalization constants $Z_\mr{S}$ and $Z_\mr{P}$, being crucial elements for the computation of renormalized quark masses based on the bare subtracted and PCAC quark mass definitions respectively, differ for finite lattice spacings. For $N_\mr{f}=3$ lattice QCD with Wilson-clover fermions and Lüscher-Weisz gluons, $Z_\mr{P}$ was determined in \cite{Campos2018}. $Z_\mr{S}$, however, is not known for this setup. Our approach is to calculate the ratio of the two, $Z_\mathrm{P}/Z_\mathrm{S}$, in which the scale dependence cancels.
Ward identities have been introduced in the past for the determination of $Z_\mathrm{P}/Z_\mathrm{S}$ in lattice QCD with Wilson-type fermions \cite{Bochicchio1985,Maiani1987}. These Ward identities have been obtained for large volumes, typically with (anti)periodic boundary conditions. Thus in numerical simulations, $Z_\mathrm{P}/Z_\mathrm{S}$ was computed for several non-zero quark masses and consequently extrapolated to the chiral limit \cite{Martinelli1993,Crisafulli1997,Bhattacharya1999,Bhattacharya2000,Bhattacharya2005c}.

Our approach is based on dedicated gauge ensembles with Schrödinger functional boundary conditions, which has two key benefits: (i) The renormalization constants are not correlated with the bare quantities they will eventually renormalize; the latter are obtained from large-volume simulations. (ii) The special boundary conditions allow simulating at, or in practice very close to the chiral limit, facilitating a massless renormalization scheme; the cutoff effects proportional to the finite mass are under control.
A sketch of the general idea and the Ward identity involved were presented at last years' lattice conference \cite{Heitger2017a} and will be elaborated in full length in a future publication \cite{Heitger2018}. 
In this status report we recall the basic elements, summarize preliminary results, and discuss sources of systematic uncertainties and cutoff effects which are currently under investigation. We note that, using a different method based on coordinate space renormalization in large volume, $Z_\mathrm{P}/Z_\mathrm{S}$ has also been computed for the same lattice action and coupling range in \cite{Korcyl2017}.

\section{Ward identities}
Our approach to $Z_\mathrm{P}/Z_\mathrm{S}$ uses the transformation property of the pseudoscalar density under small chiral rotations:
\begin{align}
\delta_\mathrm{A}^aP^b(x)=d^{abc}S^c(x)+\frac{\delta^{ab}}{N_\mathrm{f}}\bar{\psi}(x)\psi(x)\,. \label{eq:dP}
\end{align}
For $\mathfrak{su}(N_\mr{f})$ algebras with $N_\mr{f}\geq3$ the totally symmetric structure constant $d^{abc}$ is non-zero, and we can establish a relation between the scalar and the pseudoscalar density, which allows us to access the non-singlet renormalization constants. By inserting equation \ref{eq:dP} into the axial Ward identity, and employing Schrödinger functional boundary conditions, we arrive at (see \cite{Heitger2018} for details)

\begin{align}
\begin{split}
&Z_\mathrm{A}{Z_\mathrm{P}}\big[1+ab_\mathrm{A}m_\mathrm{q}+a\bar{b}_\mathrm{A}\mathrm{tr}(M)\big]\big[1+ab_\mathrm{P}m_\mathrm{q}+a\bar{b}_\mathrm{P}\mathrm{tr}(M)\big]\times \\
&\big[ f_\mathrm{AP}^{\mathrm{I},abcd}(y_0+t,y_0)-f_\mathrm{AP}^{\mathrm{I},abcd}(y_0-t,y_0)-2m\,\tilde{f}_\mathrm{PP}^{abcd}(y_0+t,y_0-t) \big]\\
=&-{Z_\mathrm{S}}\big[1+ab_\mathrm{S}m_\mathrm{q}+a\bar{b}_\mathrm{S}\mathrm{tr}(M)\big]f_\mathrm{S}^{abcd}(y_0)+\mathrm{O}(a^2)+\mathrm{O}(am)\,,
\end{split} \label{eq:wardidentity}
\end{align}
where $f_\mr{AP}$, $\tilde{f}_\mr{PP}$ and $f_\mr{S}$ are Schrödinger functional correlators with proper boundary fields $\mathcal{O}$:
\begin{align}
f_\mathrm{AP}^{\mathrm{I},abcd}(y_0+t,y_0)&=-\frac{a^6}{(N_\mr{f}^2-1)L^6}\sum_{\mathbf{x},\mathbf{y}}\langle\mathcal{O}^{\prime a}(A_\mr{I})_0^b(y_0+t,\mathbf{x})P^c(y_0,\mathbf{y})\mathcal{O}^d\rangle\,,\\
\tilde{f}_\mathrm{PP}^{\mathrm{I},abcd}(y_0+t,y_0-t)&=-\frac{a^7}{(N_\mr{f}^2-1)L^6}\sum_{x_0=y_0-t}^{y_0+t}w(x_0)\sum_{\mathbf{x},\mathbf{y}}\langle\mathcal{O}^{\prime a}P^b(x_0,\mathbf{x})P^c(y_0,\mathbf{y})\mathcal{O}^d\rangle\,,\\
f_\mathrm{S}^{\mathrm{I},abcd}(y_0)&=-\frac{a^3}{(N_\mr{f}^2-1)L^6}d^{bce}\sum_{\mathbf{y}}\langle\mathcal{O}^{\prime a}S^e(y_0,\mathbf{y})\mathcal{O}^d\rangle\,.
\end{align}

\begin{figure}
	\centering
	\includegraphics[width=0.5\linewidth]{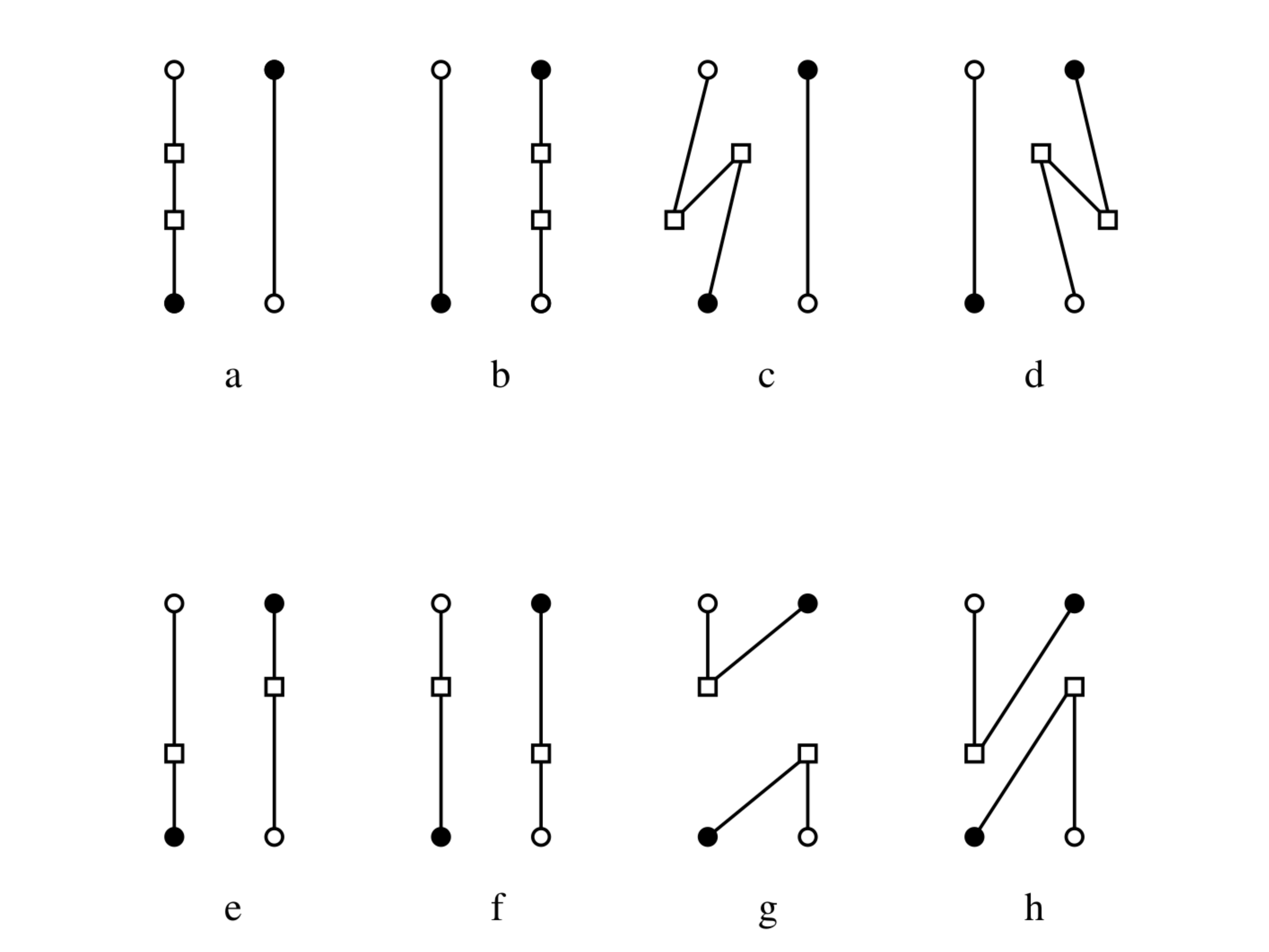}
	\caption{Graphical representation of the Wick contractions contributing to $f_{\Gamma\tilde{\Gamma}}$, taken from \cite{Luscher1996}.}
	\label{fig:contractions}
\end{figure}

Equation \ref{eq:wardidentity} can be solved for $Z_\mathrm{P}/Z_\mathrm{S}$, using our knowledge of $Z_\mr{A}$ \cite{Brida2018b,Bulava2016} and neglecting contributions multiplying the $b_\mr{X}$ and $\bar{b}_\mr{X}$ coefficients, as they vanish in the chiral limit (i.e., $m_\mr{q}=0$ in the valence quark sector and $\mr{tr}(M)=0$ in the sea). The four-point functions can be calculated as the sum of the eight Wick contractions graphically depicted in figure \ref{fig:contractions}. Each of the eight terms is proportional to the trace of the product of four SU$(N_\mr{f})$ generators $T^a, T^b, T^c$ and $T^d$. As the flavour indices $a,b,c$ and $d$ are not uniquely determined, we can make several choices which result in different Ward identities. In this status report we will focus on two specific choices which have no contributions from the disconnected diagrams labeled g and h in figure \ref{fig:contractions} and are thus less prone to statistical noise. First, our preferred flavour choice is $a=2,b=5,c=6$ and $d=8$, from now on labeled WI($2568$). Second, as a crosscheck we employ the difference of WI($8383$)$-$WI($4141$) where the disconnected contributions cancel. The $Z_\mathrm{P}/Z_\mathrm{S}$ results from WI($2568$) and the difference WI($8383$)$-$WI($4141$) are expected to differ only by $\mathrm{O}(a^2)$. The Ward identities are in principle valid for any choice of $y_0$ and $t$, but as we expect contamination close to the temporal borders of the lattice, we decided to impose the renormalization conditions at $y_0=T/2$ and $t=T/4$.

\vspace{-7pt}
\section{Numerical details}

Our simulations are based on gauge ensembles with Schrödinger functional boundary conditions, also used in previous studies \cite{Heitger2017a,Bulava2015b,Bulava2016}. We use Wilson-clover quarks and tree-level Symanzik improved gluons. The ensembles lie on an approximate line of constant physics with $L\approx 1.1\,$fm. In table \ref{tab-simulation_parameters} we present the simulation parameters for the individual ensembles. We simulate at five different bare couplings, covering the range of the $N_\mr{f}=2+1$ CLS ensembles \cite{Bruno2014b}, which are currently being used for the computation of light quark masses \cite{Jonna}. For each coupling we investigate several values for $\kappa$, corresponding to PCAC masses $|am|\leq 0.014$, to be able to extra- or interpolate to the chiral limit. For now we have one ensemble, labeled B2k1, which violates the constant physics condition and can be used for checking systematic effects. At the smaller lattice spacings, our simulations suffer from critical slowing down in the topological charge. Previous analyses have shown that the influence of this issue on renormalization constants is negligible. In this preliminary analysis we project all observables to the $Q=0$ sector to circumvent this problem and leave a comparison with a full analysis for the final publication \cite{Heitger2018}.

\begin{table}[thb]
		\footnotesize
		\centering
		
		\begin{tabular}{cllrrc}
			\toprule
			$L^3\times T/a^4$  & $\beta$ & $\kappa$ & \#REP & \#MDU & ID \\
			\midrule
			$12^3\times17$ & 3.3 & 0.13652 & 20 & 10240 & A1k1 \\
			
			&  & 0.13660 & 10 & 13672 & A1k2 \\
			&  & 0.13648 & 5 & 6876 & A1k3 \\
			\midrule
			$14^3\times21$ & 3.414 & 0.13690 & 32 & 25600 & E1k1 \\
			&  & 0.13695 & 48 & 38400 & E1k2 \\
			\midrule
			$16^3\times23$ & 3.512 & 0.13700 & 2 & 20480 & B1k1 \\
			&  & 0.13703 & 1 & 8192 & B1k2 \\
			&  & 0.13710 & 3 & 22528 & B1k3 \\
			\midrule
			$16^3\times23$ & 3.47 & 0.13700 & 3 & 29560 & B2k1 \\
			
			\midrule
			$20^3\times29$ & 3.676 & 0.13700 & 4 & 15232 & C1k2 \\
			&  & 0.13719 & 4 & 15472 & C1k3 \\
			\midrule
			$24^3\times35$ & 3.810 & 0.13712 & 6 & 10272 & D1k1 \\
			&  & 0.13701& 3 & 5672 & D1k2 \\
			&  & 0.137033& 7 & 6488 & D1k4 \\
			
			\bottomrule
			
		\end{tabular}		
		\caption{Summary of simulation parameters of the gauge configuration ensembles used in this study, as well as
			the number of (statistically independent) replica (\#REP) per ensemble ‘ID’ and their total number of molecular dynamics
			units (\#MDU).}
		\label{tab-simulation_parameters}
\end{table}

\section{Results}
\begin{figure}[!htb]
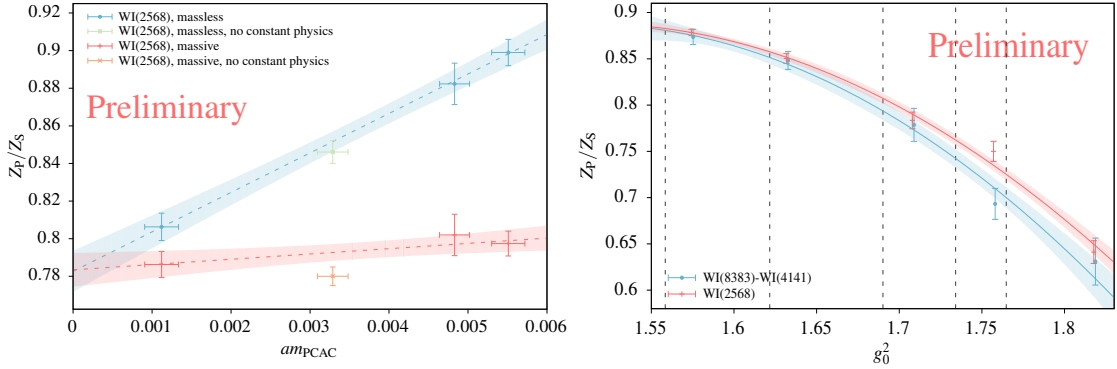

	\centering
	\begin{minipage}{.49\textwidth}
		\centering
		\graphicspath{{figures/}}
		\resizebox{\textwidth}{!}{\input{figures/B_0_m_chiral}}
	\end{minipage}
	\begin{minipage}{0.49\textwidth}
		\centering
		\vspace{6.8pt}
		\graphicspath{{figures/}}
		\resizebox{\textwidth}{0.71\textwidth}{\input{figures/zszp_Q0}}
	\end{minipage}
	\caption{Left: Preliminary chiral extrapolation of $Z_\mathrm{P}/Z_\mathrm{S}$ derived from WI($2568$) without and with mass term for $g_0^2=1.7084$. Data points from ensemble B2k1 violate the constant physics condition and are shown for comparison. Right: Preliminary results for $Z_\mathrm{P}/Z_\mathrm{S}$ from WI($8383$)-WI($4141$) and WI($2568$) with interpolating Padé fits. Dashed lines indicate the bare couplings used in CLS simulations.}
	\label{fig:double_fig}
\end{figure}

We evaluate the correlation functions required for \ref{eq:wardidentity} on the gauge ensembles presented in table \ref{tab-simulation_parameters}. We use the axial vector current improvement factor $c_\mathrm{A}$ from \cite{Bulava2015b} and the axial vector renormalization constant from \cite{Brida2018b}.
Our error analysis is based on the $\Gamma$-method, taking into account slow modes in the simulation \cite{Wolff2003,Schaefer2010}.
The multiple values of $\kappa$ for each bare coupling $g_0^2$ enable us to reliably impose a renormalization condition at zero quark mass, where the leading cutoff effects of $\mathrm{O}(am)$ vanish.
In figure \ref{fig:double_fig} the chiral extrapolation is illustrated for $g_0^2=1.7084$. The mass-dependent improvement factors denoted $b_\mathrm{X}$ and $\tilde{b}_\mr{X}$ in \ref{eq:wardidentity} are neglected, as their contribution vanishes in the chiral limit and their inclusion would only result in a different slope. 

The deviation from the chiral limit is estimated by the PCAC mass with improved definition of the lattice derivative, which we average over the central third of the temporal extent of the lattice.  We compare two different $Z_\mathrm{P}/Z_\mathrm{S}$ determinations, both based on WI($2568$). The data points labeled ‘massless’ are determined by neglecting the mass term in equation \ref{eq:wardidentity}, which arises as a contact term, while points labeled ‘massive’ are based on the full Ward identity. Both conditions agree in the linear extrapolation to the chiral limit. The inclusion of the mass term allows for a smoother extrapolation with a smaller uncertainty.

As mentioned earlier, all our gauge ensembles are tuned to lie on a line of constants physics. This tuning was done via the universal 2-loop $\beta$-function as explained in \cite{Bulava2015b}. After the final scale setting \cite{Bruno2016b} it turned out, that the resulting variation of the box size is of the order $10\,\%$. A variation of the physical box size is expected to effect the renormalization constants at order $(a/L)^2$ \cite{Luscher1996} and could possibly have a similar magnitude as the statistical uncertainty. Since $Z_\mathrm{P}/Z_\mathrm{S}$ has not been determined in the Schrödinger functional before, it is crucial to further investigate this issue.

For now, we are only able to estimate the violation of the constant physics condition with the help of ensemble B2k1, where we simulated an additional bare coupling at $L/a=16$ and one value of $\kappa$.
The respective data points are depicted in the left part of figure \ref{fig:double_fig}, labeled ‘no constant physics’. They align fairly well with the data points from ensembles B1 within the statistical uncertainty, which gives the impression that the aforementioned issue is subleading. Unfortunately, two contrary effects obstruct a definite conclusion: While the slightly higher bare coupling is expected to shift $Z_\mathrm{P}/Z_\mathrm{S}$ to a lower value, the finite positive mass does the opposite. The influence of the varying box size is therefore not decisively clear.
To settle this issue, we plan to perform further simulations at $\beta=3.676$ and vary the physical extent from $L/a=20$ to $16$, $12$ and $8$ with two values of $\kappa$ each. This will enable us to judge how a moderate to stark variation of the physical extent of the lattice influences our observable in the chiral limit.

Our preliminary results for $Z_\mr{P}/Z_\mr{S}$ are presented in the right part of figure \ref{fig:double_fig}, the given uncertainties are the statistical error from the linear fit to the chiral limit and a systematic component estimated by the difference of the massless and the massive renormalization conditions, both added in quadrature. Effects from the violation of the line of constant physics are not included yet. For orientation, the bare couplings of the $N_\mr{f}=2+1$ CLS ensembles are indicated by dashed lines. The data points derived from both $Z_\mathrm{P}/Z_\mathrm{S}$ determinations align nicely and seem to approach each other towards smaller bare couplings. We observe that the absolute statistical uncertainties grow with increasing coupling. This makes it hard to discern whether the difference between the two determinations are actually of leading $\mathrm{O}(a^2)$, as it should be in the $\mathrm{O}(a)$ improved theory. The data points at $g_0^2=1.8181$ seem a little off, as the two points agree within statistical error, while we expect a larger deviation. However, we want to remind the reader that CLS also considered this inverse coupling at an initial stage of their simulations, but later discarded these ensembles, because large cutoff effects were observed \cite{Bruno2014b}. In fact a similar effect could come to play in our project as well, and one could think about giving the data points a lower weight for the final interpolation formula as it is irrelevant for the CLS range and just used to stabilize the fit. To settle this issue, we plan to increase the number of gauge configurations for the largest bare coupling by a factor of $8$.

Another source of cutoff effects we have not investigated yet, are the insertion times of the operators in equation \ref{eq:wardidentity}. While the Ward identities are valid for any choice of $y_0$ and $t_1$ by definition, boundary effects and contact terms can lead to additional cutoff effects. To get an idea of the size of these effects, we plan to vary the insertion times systematically for one bare coupling.

\section{Outlook}
In this report we focus on two possible choices of flavour indices. There are several other choices under investigation, in some of which the disconnected quark diagrams contribute. Each choice of flavour indices, insertion times for the Ward identities, and plateau region for the PCAC mass leads to a determination which differs from others by $\mathrm{O}(a^2)$ terms. By comparing the different $Z_\mr{P}/Z_\mr{S}$ results and by increasing the statistical precision at the coarsest lattice spacing under investigation, we will be able to estimate these cutoff effects more accurately.
The only source of systematic uncertainty that has not yet been accounted for in our analysis is the violation of the constant physics condition. We are currently examining its effects along the lines described above. As a final crosscheck we will be able to compare our results with the ongoing determination of $Z=Z_\mathrm{P}/(Z_\mathrm{S}Z_\mathrm{A})$, reported in \cite{DeDivitiis2017,DeDivitiis2018}, combining it with the axial vector renormalization constant. 

\section*{Acknowledgments}
We thank S.\ Sint, C.\ Wittemeier and S.\ Kuberski  for helpful discussions. A.\ Vladikas wishes to thank the Particle Physics Theory Group at the WWU Münster for hospitality. This work is supported by the Deutsche Forschungsgemeinschaft (DFG) through the Research Training Group \textit{“GRK 2149: Strong and Weak Interactions – from Hadrons to Dark Matter”} (J. H. and F. J.). We acknowledge the computer resources provided by the \textit{Zentrum für Informationsverarbeitung} of the University of Münster (PALMA \& PALMA II HPC clusters) and thank its staff for support.

\providecommand{\href}[2]{#2}\begingroup\raggedright\endgroup

\end{document}